\documentclass{PoS}

\title{Exclusive pion-induced Drell-Yan process at J-PARC for accessing the nucleon GPDs and soft nonfactorizable mechanism}

 \ShortTitle{Exclusive pion-induced Drell-Yan process at J-PARC 
 for accessing the nucleon GPDs 
 }

 \author{\speaker{Kazuhiro Tanaka}%
\\
        Department of Physics, Juntendo University, Inzai,
  Chiba 270-1695, Japan and\\
J-PARC Branch, KEK Theory Center, Institute of
  Particle and Nuclear Studies, KEK, 203-1, Shirakata, Tokai, Ibaraki,
  319-1106, Japan\\
        E-mail: \email{kztanaka@juntendo.ac.jp}}

\abstract{Generalized parton distributions (GPDs) encoding multidimensional information of
hadron partonic structure appear as the building blocks in a factorized description of
hard exclusive reactions. The nucleon GPDs have been accessed by deeply virtual
Compton scattering and deeply virtual meson production with lepton beam. A
complementary probe with hadron beam is 
shown to be  
the exclusive pion-induced Drell-Yan process, $\pi^- p \to \mu^+ \mu^- n$,
as demonstrated by recent theoretical advances on describing this process in terms of QCD factorization as the
partonic subprocess convoluted with the nucleon GPDs and the pion distribution
amplitudes,  
and by the feasibility study for its measurement
via a spectrometer at the
High Momentum Beamline being constructed at J-PARC in Japan.
We also discuss
the possible soft partonic mechanisms beyond the QCD factorization framework, 
and present an estimate of the soft mechanisms at J-PARC kinematics, making use of dispersion
relations and quark-hadron duality. 
Realization of the measurement of the exclusive pion-induced Drell-Yan process at J-PARC will provide a new test of QCD descriptions of a novel class of hard exclusive reactions, and also offer the possibility of experimentally accessing nucleon GPDs at large timelike virtuality. 
}

\FullConference{XXV International Workshop on Deep-Inelastic Scattering and Related Subjects\\
		3-7 April 2017\\
		University of Birmingham, UK}

\usepackage{wrapfig}
\usepackage{graphics}
\usepackage[dvips]{epsfig}
\newcommand{\bc}{\begin{center}}
\newcommand{\ec}{\end{center}}
\usepackage{bm}

\usepackage{color}

\begin{document}

\section{Introduction}
\label{sec:1}
\begin{wrapfigure}{l}{6.2cm}
\vspace{-0.8cm}
{\includegraphics[width=0.45\textwidth]{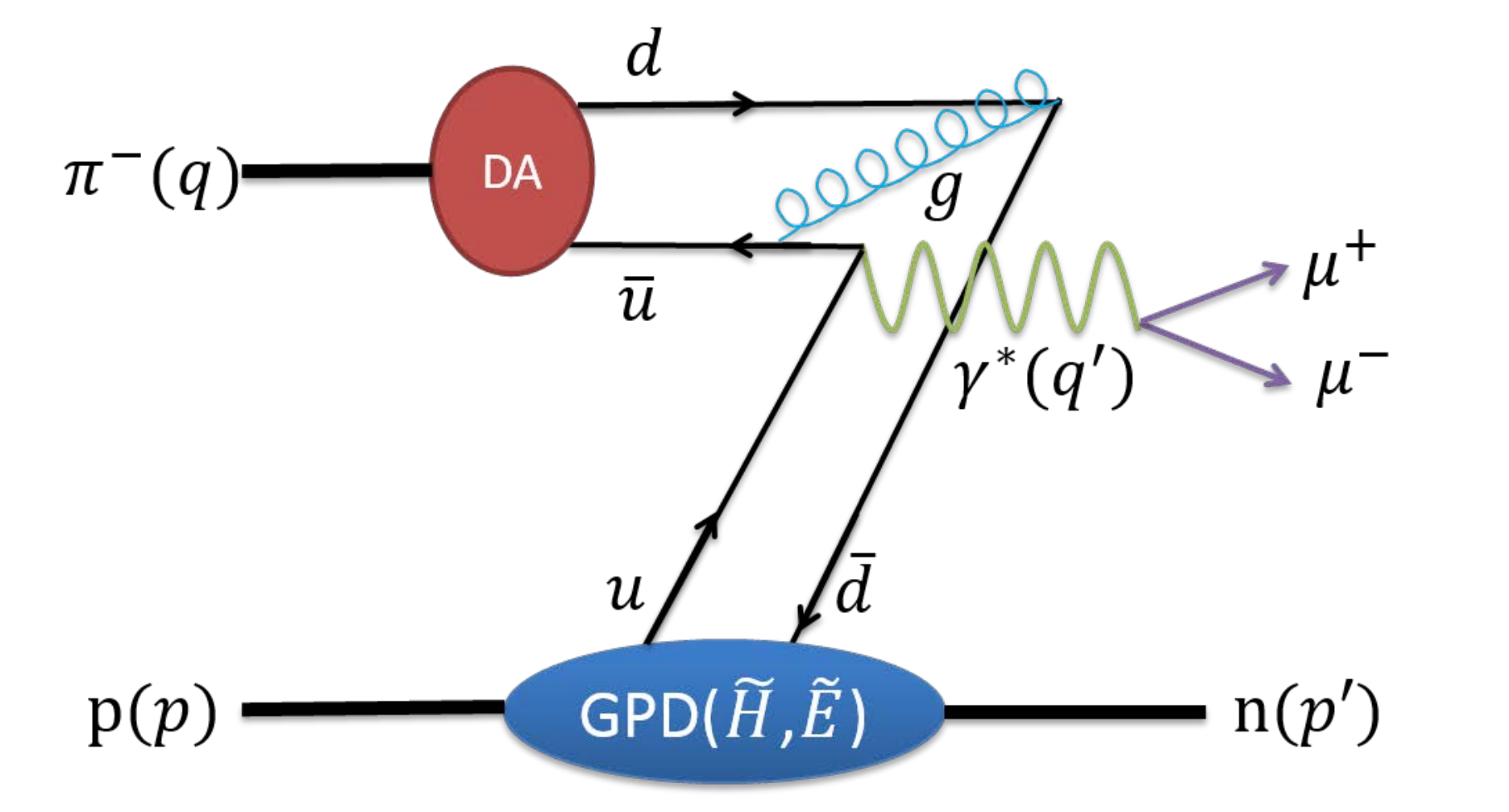}
\label{fig:edycross2}}
\caption{\label{fig1}Exclusive pion-induced Drell-Yan process for the forward
dimuon production.}
\vspace{-0.3cm}
\end{wrapfigure}
We consider the pion-induced dimuon production. Summing the absolute square of the corresponding amplitudes, 
$\pi N\to q\bar q X \to\gamma^*  X\to \mu^+\mu^- X$, over the accompanying hadronic final state $X$, 
we obtain the inclusive Drell-Yan cross section. The leading contribution comes from the transversely-polarized virtual photon $\gamma^*$, as a consequence of the helicity conservation
in the annihilation of the on-shall, massless quark and antiquark associated with the relevant partonic subprocess $q \bar q \to \gamma^*$.
%(See e.g. \cite{Peng:2015spa} and references therein.) 
On the other hand, the dimuon angular distribution for the production in the forward region is known to obey the pattern associated with the longitudinally-polarized virtual photon, which can be produced 
by the annihilation of off-shell quark $q$ or antiquark $\bar q$.
The relevant off-shellness 
may be 
caused by perturbative gluon exchange between the quark and antiquark originating from the pion,
and this type of mechanism with the gluon exchange plays important role for the forward production. 
Now, the spectator quark (or antiquark) originating from the pion may be absorbed by the target nucleon, 
giving rise to the exclusive final state, $\mu^+\mu^- N$, as represented  
in Fig.~\ref{fig1}. 
This is the exclusive Drell-Yan process and this type of diagrams gives important contributions 
for the dimuon production in the forward region
of the exclusive Drell-Yan process~\cite{Berger:2001zn}.

We discuss a recent cross section estimate of the exclusive Drell-Yan process 
as a report of our recent paper~\cite{Sawada:2016mao}.
With the 
High Momentum Beamline being constructed at J-PARC,
%high-momentum beam line at J-PARC, 
the secondary pion beam with moderately high energies $\sim 15$-$20$~GeV 
is best suited to study meson-induced hard exclusive processes 
like exclusive Drell-Yan process.
We mention the feasibility study~\cite{Sawada:2016mao} for measuring the exclusive
pion-induced Drell-Yan process with the E50 spectrometer at J-PARC.
We also discuss non-factorizable mechanism, beyond QCD factorization, 
in exclusive Drell-Yan process and give its first estimate 
using the light-cone QCD sum rules~\cite{Tanaka:2017rym,tanaka}.

\section{QCD factorization formula at the leading order (LO)}
\label{sec:2}
We consider the exclusive Drell-Yan production, $\pi^- p \to \gamma^* n \to \mu^+ \mu^- n$, 
%as in the right diagram of Fig.~\ref{fig1}, 
in particular, with
%We consider 
the production of $\gamma^*$ in the forward region corresponding to the small invariant-momentum-transfer,
$t =\Delta^2$ with $\Delta\equiv q-q'$, where $q$ and $q'$ are the momenta of the initial pion and the produced $\gamma^*$. 
In this case, as mentioned 
%in the right diagram of Fig.~\ref{fig1},
in Sec.~\ref{sec:1}, 
the complete annihilation of quark as well as antiquark from the pion,
as in Fig.~\ref{fig1},
plays important role. Here, the relevant amplitude is expressed as the convolution of the corresponding 
partonic (short distance) annihilation processes with the two separate parts of long-distance nature, associated with the pion and the nucleon, respectively:
the upper long-distance part denotes the pion distribution amplitude (DA), whose information can be obtained from, e.g., $\gamma \gamma^* \to \pi^0$ process at Belle and Babar,
while the lower long-distance part denotes the generalized parton distribution functions (GPDs) as an 
off-forward nucleon matrix element, whose forward limit reduces to the usual helicity distribution, $\Delta q(x)$.
Here, the GPDs are defined
as the off-forward matrix element of the bilocal light-cone operators of the type, 
$\langle n(p')|\bar q(-y^-/2)\cdots q(y^-/2)|p(p)\rangle$, with $y^+ = \vec y_\perp =0$ ($y^\mu y_\mu=0$),  and are the functions of
the relevant invariants, the (average) 
light-cone momentum fraction $x$ and the skewness $\xi$ ($=(p-p\,')^+/(p+p\,')^+$), 
as well as $t$ (see Sec.~II in \cite{Sawada:2016mao}).   
Decomposing $\langle n(p')|\bar q(-y^-/2)\gamma^+ q(y^-/2)|p(p)\rangle$
%the above type of matrix elements
into the independent Lorentz structures,
we obtain the familiar proton GPDs, $H^q, E^q$,
relevant to the Ji sum rule for quark's angular momentum contribution to the proton spin, 
$J^q = \int_{-1}^{1} dx  x  [H^q (x,\xi,0)$ $+E^q (x,\xi,0) ] / 2$, 
and also $\tilde H^q, \tilde E^q$ for the case with the additional $\gamma_5$, as ($P\equiv (p+p')/2$),
\begin{equation}
%&&
 2  P^+\int  \frac{d y^-}{4\pi}e^{i x P^+ y^-}
 \langle p' | 
 \bar{q}(-\frac{y^-}{2}) \gamma^+ \gamma_5 q(\frac{y^-}{2}) 
 | p \rangle
= \bar{u} (p') 
 \Bigl [ \tilde{H}^q (x,\xi,t) \gamma^+ \gamma_5
     + \tilde{E}^q (x,\xi,t)  \frac{\gamma_5 \Delta^+}{2 \, m_N}
 \Bigr] u (p)\ ,
\label{eqn:gpd-p}
\end{equation}
%we obtain the familiar proton GPDs, $H^q, E^q$, 
for each quark flavor $q$;
here, 
$| p^{(\prime)} \rangle \equiv | p (p^{(\prime)})\rangle$,
$u(p^{(\prime)})$ denotes the proton
spinor with momentum $p^{(\prime)}$ and mass $m_N$, 
and we do not show the gauge-link operator between
two quark fields.
Those GPDs, $H, E, \tilde H$, and $\tilde E$, are measured by the deeply virtual Compton scattering (DVCS) 
corresponding to $\gamma^* p \to \gamma p$ process in the experiments at JLab, HERMES, COMPASS, etc., 
and also by the deeply virtual meson production (DVMP).
For the case of the deeply virtual pion production, $\gamma^* p \to \pi N$,
the pseudoscalar nature of the pion allows us to probe the GPDs $\tilde H^q$ and $\tilde E^q$ solely, 
associated with $\gamma_5$ as in (\ref{eqn:gpd-p}).
Interchanging the initial 
%virtual photon 
$\gamma^*$ and the final pion in the deeply virtual pion production and making the $\gamma^*$ timelike, 
we obtain the exclusive Drell-Yan process.
This demonstrates that the exclusive Drell-Yan process at the J-PARC allows us to probe $\tilde H^q$ and $\tilde E^q$ solely and plays a complementary role compared with the deeply virtual pion production at, e.g., JLab. The kinematical region accessible by the exclusive Drell-Yan process at the J-PARC is also complementary to those accessible by the GPD measurements 
by the various other experiments~\cite{Sawada:2016mao}.

\begin{wrapfigure}{l}{7cm}
\vspace{-0.6cm}
\includegraphics[width=0.5\textwidth]{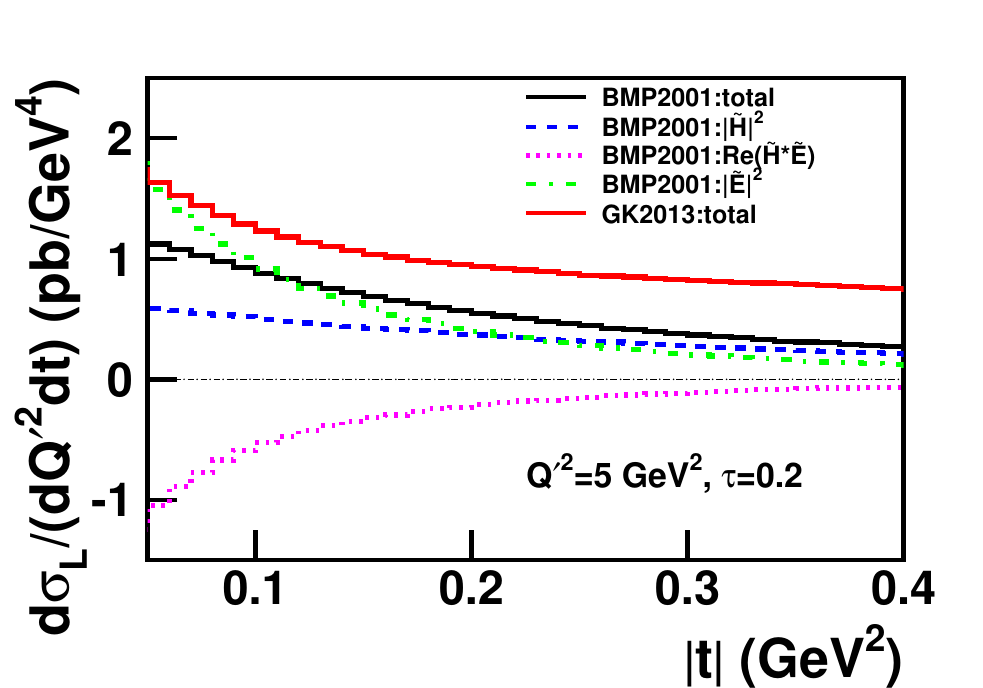}
\caption
[\protect{}] {Cross section~(\ref{eq_dcross}) 
of
$\pi^- p \to \gamma^* n$ with $Q'^2=5$~GeV$^2$
as a function of $|t|$ for $\tau = 0.2$~\cite{Sawada:2016mao}. 
    }
\label{fig:edycrossPLB}
\vspace{-0.2cm}
\end{wrapfigure}

In Fig.~\ref{fig1},
the ``hard'' gluon exchange ensures that the vertices in the partonic subprocess are separated by short distances;
thus, the diagrams associated with this type of gluon exchange obey the QCD factorization into 
the corresponding short-distance partonic subprocess and  
the long-distance parts, 
the pion DA and the nucleon GPD. 
Indeed, Fig.~\ref{fig1} and similar diagrams 
%at the same order in 
of order $\alpha_s$
%The diagrams associated with this type of gluon exchange
give 
the leading order (LO) in the  factorization formula for exclusive Drell-Yan process.
The first estimate 
%of the exclusive Drell-Yan cross section 
using this LO factorization formula was performed 
by Berger, Diehl and Pire~\cite{Berger:2001zn}. 
The corresponding cross section at the large $Q'^2\equiv q'^2$ scaling limit with the fixed $\tau \equiv Q'^2/(2 p\cdot q)$ is 
\begin{eqnarray}
\frac{d\sigma_L}{dt dQ'^2}
&=& \frac{4\pi \alpha_{\rm em}^2}{27}\frac{\tau^2}{Q'^8} f_\pi^2\, \Bigl[ (1-\xi^2) \left|\tilde{\cal H}^{du}
%(\tilde{x},\xi,t)
\right|^2
- 2 \xi^2  
{\rm Re} \bigl( \tilde{\cal H}^{du*}
%(\tilde{x},\xi,t)^* 
\tilde{\cal E}^{du}
%(\tilde{x},\xi,t) 
\bigr)
   -   \frac{\xi^2t}{4 m_N^2}\left|\tilde{\cal E}^{du}
%(\tilde{x},\xi,t)
   \right|^2 \Bigr]\ ,
\label{eq_dcross}\\
\tilde{\cal H}^{du}
%(\tilde{x},\xi,t)
=&&
\!\!\!\!\!\!\!\!\!\!\!\!
\frac{8\alpha_s}{3} \int_{-1}^1
  dz\, \frac{\phi_\pi(z)}{1-z^2} \int_{-1}^1 dx
  \Bigl( 
 \frac{e_u}{\xi-x+ i\epsilon}-\frac{e_d}{\xi+x+ i\epsilon}
  \Bigr)  \bigl[ \tilde{H}^{d}(x,\xi,t) - \tilde{H}^{u}(x,\xi,t)
  \bigr]\ ,
\label{eq_Hdu}
\end{eqnarray}
where $f_\pi$ is the pion decay constant,
%($\tilde x=-\xi$, 
$e_q$ is the quark's electric charge,
and $\tilde{\cal H}^{du}$ is a function of $\xi$ and $t$ 
as the 
%absolute square of a 
convolution of the hard part, the pion DA $\phi_\pi(z)$ of leading twist, and the GPD $\tilde H^q$,
%and the absolute square of 
while $\tilde{\cal E}^{du}$  denotes the similar convolution 
with $\tilde H^q$ replaced by the GPD $\tilde E^q$.
%arises as its absolute square and its interference with $\tilde{\cal H}^{du}$.
In (\ref{eq_Hdu}), the $p\to n$ transition GPDs arising in Fig.~\ref{fig1} 
is expressed by the proton GPDs of (\ref{eqn:gpd-p}) using isospin invariance relations~\cite{Berger:2001zn,Sawada:2016mao}. 
The subscript ``$L$'' in (\ref{eq_dcross}) indicates that this cross section 
is obtained at 
the leading twist, associated with 
the production of the longitudinally-polarized $\gamma^*$.
%For a treatment of 
(The production of the transversely-polarized 
%with the production of the transversely-polarized 
$\gamma^*$ requiers higher twist effects,
see \cite{Goloskokov:2015zsa}.)
%\newline
With $Q'^2=5$~GeV$^2$ for the mass of the produced dimuon, 
the cross section~(\ref{eq_dcross}) is plotted in Fig.~\ref{fig:edycrossPLB}.
The results of \cite{Berger:2001zn} are labeled  as
``BMP2001'', where a model of the nucleon GPDs $\tilde H$ and $\tilde E$, based on the double distributions, and the asymptotic pion DA for $\phi_\pi$ are used. 
The cross section is of the pb level.
Recently~\cite{Sawada:2016mao}, we have updated the estimate of the corresponding cross section
and obtained the results labeled as ``GK2013'' in Fig.~\ref{fig:edycrossPLB}: we have used a recent parameterization for the GPDs $\tilde H$ and $\tilde E$, determined by comparing with the HERMES data for $\pi^+$ electroproduction, as well as to the pion DA with the pre-asymptotic corrections for $\phi_\pi$. 
The updated cross section is enhanced compared with the previous result. For more detail, we refer the readers to \cite{Sawada:2016mao}.

Using this updated estimate of (\ref{eq_dcross}) as an input, 
we have performed the Monte Carlo simulation and the feasibility study~\cite{Sawada:2016mao} for measuring 
$\pi^- p \to \mu^+ \mu^- n$ at J-PARC. 
We are able to obtain Monte Carlo simulation signals of the dimuon mass spectra for 
%the values of the pion beam momentum corresponding to 
the secondary pion beam with the J-PARC high-momentum beam line,
%These results are obtained 
assuming a minimal extension of the E50 spectrometer 
%for the charmed-baryon spectroscopy experiment 
at J-PARC
(see Fig.~9 in \cite{Sawada:2016mao}). 
Our results of the Monte Carlo simulated missing-mass $M_X$ spectra 
demonstrate that 
the exclusive Drell-Yan signal is well-separated from the inclusive as well as other signals, see Fig.~14 in \cite{Sawada:2016mao}. 
%It is also demonstrated~\cite{Sawada:2016mao} that 
Also, the accuracy expected for the corresponding J-PARC data 
%of the exclusive Drell-Yan process 
will allow us to 
distinguish the typical parameterizations for the GPDs.
For further detail, see \cite{Sawada:2016mao}. 

\section{Nonfactorizable mechanism}
\label{sec:3}
To calculate the convolution implied by 
Fig.~\ref{fig1},
corresponding to the LO factorization for exclusive Drell-Yan process, we integrate the corresponding amplitudes over the momentum associated with the gluon propagator. When the gluon momentum becomes small compared with $\Lambda_{\rm QCD}$, such soft and nonperturbative degrees of freedom should be separated into the long-distance parts in the spirit of  QCD factorization.
Absorbing the soft nonperturbative gluon propagator into either the pion DA or the nucleon GPDs 
leads to the ``tree'' diagrams, which are obtained formally by removing the gluon propagator from 
%the lower diagram of 
Fig.~\ref{fig1}.
%and the diagrams of similar type. 
Thus, the tree diagrams correspond to the lower order in $\alpha_s$ than the LO in the QCD factorization framework
and physically represent the ``Feynman mechanism'':
the antiquark (quark) carrying almost all pion-momentum annihilates with the quark (antiquark) carrying almost all momentum-transfer from the nucleon, 
to produce $\gamma^*$,
while the ``wee'' parton is directly transferred between the pion and the nucleon. 
The corresponding partonic process is not ensured to be of short-distance, 
and thus this diagram is not factorizable into the short- and long-distance parts. 
Moreover, we do not have a boundary to separate the pion and the nucleon wave functions
because they are directly connected by the soft parton line;
thus, the nonperturbative function arising in the tree diagrams is also nonfactorizable between those hadrons.

%It is worth noting that 
A similar soft nonfactorizable mechanism is known to play an important role 
in the QCD description of the pion electromagnetic form factor.
In addition to the QCD factorization formula 
associated with a hard-gluon exchange in the partonic subprocess,
nonfactorizable mechanisms corresponding to the ``tree'' partonic-process without gluon exchange,
where the two pions are connected by a soft paton transfered direcly between them,
are indispensable for reproducing empirical behaviors of the form factor,
%in QCD calculation, 
especially for moderate momentum-transfer region~\cite{Braun:1999uj}.

To perform QCD calculation of  the nonfactorizable mechanism in the exclusive Drell-Yan process~\cite{Tanaka:2017rym,tanaka},
% for the exclusive Drell-Yan process in a largely model-independent way, 
we first make the external leg of the initial pion {\it off-shell}, and replace the corresponding pion wave function by the axial vector vertex, to which the pion can couple.
This procedure leads to a description using the two-point correlator,
${\cal T}_{\mu\nu}=i\int  {d^4}x{\mkern 1mu} {e^{iq' \cdot x}}\langle n(p')|{\rm{T }}j_\mu ^5(0)j_\nu ^{{\rm{em}}}(x)|p(p)\rangle$,
with $q=q'+p'-p$ and  $q^2 \neq m_\pi^2$,
corresponding to the off-forward virtual Compton amplitude with one of the electromagnetic 
currents, $j_\nu ^{{\rm{em}}} = {e_u}\bar u{\gamma _\nu }u + {e_d}\bar d{\gamma _\nu }d$, replaced by the axial vector 
current, $j_\mu ^5 = \bar d{\gamma _\mu }{\gamma _5}u$.
For deeply virtual region, $| q^2|,| q'^2 | \gg \Lambda _{{\rm{QCD}}}^2$,
the correlator ${\cal T}_{\mu\nu}$ can be systematically treated by the operator product expansion (OPE),
and 
%is is straightforward to see~\cite{tanaka} that 
the corresponding long-distance contribution 
%of the corresponding OPE 
can be expressed by the nucleon GPDs $\tilde H$ and $\tilde E$, which are the same GPDs as appeared in 
%the LO QCD factorization formula~
(\ref{eq_dcross}).

We may also write down the dispersion relation for ${\cal T}_{\mu\nu}$ with respect to its dependence on $q^2$,
and it can be shown that the residue at the pion pole, $q^2=m_\pi^2$, 
for this dispersion relation corresponds to the exclusive Drell-Yan amplitude
associated with the on-shell pion leg. 
The corresponding residue may be determined 
from the behavior of the OPE for the off-forward deeply virtual amplitude ${\cal T}_{\mu\nu}$;
when we perform the OPE at the tree level, the result should give the nonfactorizable mechanism due to the tree diagrams.
For an efficient matching between the OPE and dispersion relation to determine the relevant pole residue,  
%To extract the soft nonfactorizable amplitude with the on-shell external pion from the behaviors of this OPE, 
we rely on quark-hadron duality to deal with the unwanted higher resonance contributions arising in the dispersion relation.
This procedure yields the soft nonfactorizable amplitude for  $q^2=m_\pi^2 \rightarrow 0$
%in the present kinematics 
as~\cite{Tanaka:2017rym,tanaka},
\begin{equation}
\langle n|j_\nu ^{{\rm{em}}}|\pi^- p\rangle=-g_\nu ^ -\frac{2i}{{{f_\pi }}}\int_\xi ^{{x_0}} {dx}
e^{ - \frac{(x - \xi)Q'^2}{(x + \xi)M_B^2}}
\left[ {e_u}{{\tilde H}^{du}}(x,\xi ,t) 
-
{e_d}{{\tilde H}^{du}}( - x,\xi ,t)\right]
\bar u(p'){\gamma ^ + }{\gamma _5}u(p )  
 +  \cdots\ ,
\label{sna}
\end{equation}
in terms of the proton GPDs, ${\tilde H^{du}}(x,\xi ,t) = {\tilde H^u}(x,\xi ,t) - {\tilde H^d}(x,\xi ,t)$,
%and the  coefficient functions arising in the OPE, 
and 
we have also the similar term associated with  
${\tilde E^{du}}(x,\xi ,t) = {\tilde E^u}(x,\xi ,t) - {\tilde E^d}(x,\xi ,t)$, 
as well as the terms arising from higher-twist
corrections to the OPE for ${\cal T}_{\mu\nu}$,
in the ellipses.
Here, $x_0$ is related to the threshold parameter $q_{\rm th}^2$, from which the continuum contribution in the dispersion-relation
integral for ${\cal T}_{\mu\nu}$
starts, as the approximation for the higher resonance contributions invoking quark-hadron duality. 
We note that the factor $g_\nu^-$ in (\ref{sna}) indicates the longitudinal polarization of
the produced $\gamma^*$.

\begin{figure}[hbtp]
%\begin{center}
\centering
%\subfigure[]
%{
\includegraphics[width=0.43\textwidth]{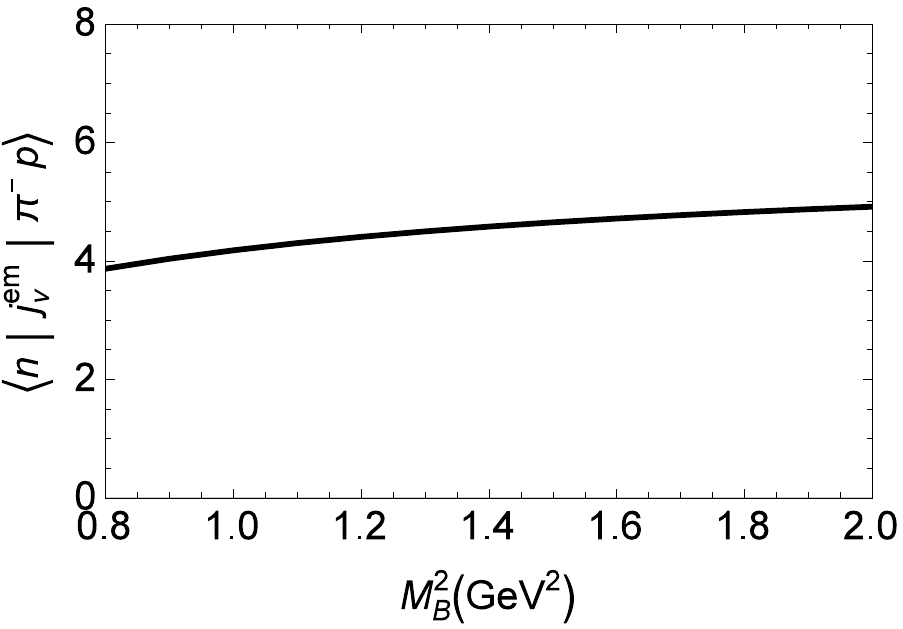}
%\label{fig:edycrossPLB_1}}
%\subfigure[]
%{
\hspace{0.7cm}
\includegraphics[width=0.45\textwidth]{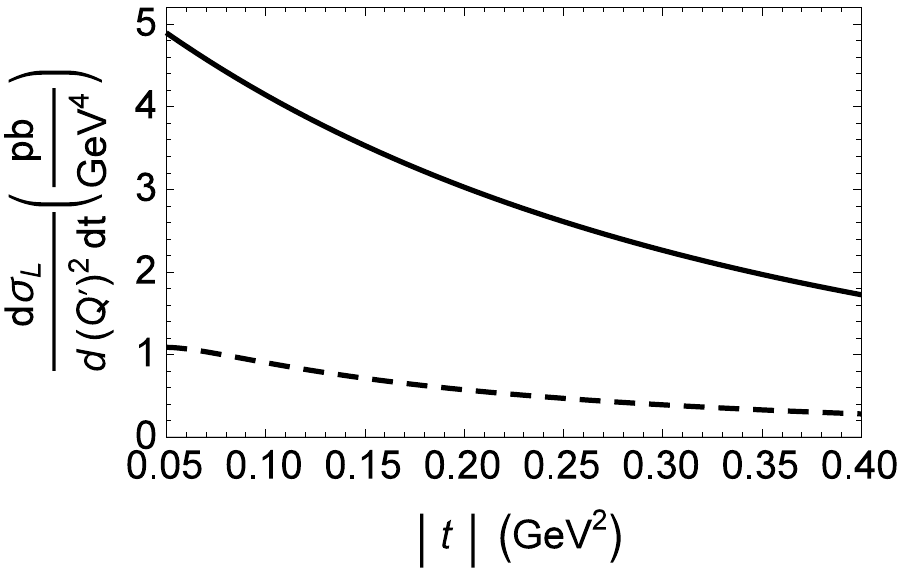}
%\label{fig:edycrossPLB_2}}
\caption
[\protect{}] {Light-cone sum rule~(\ref{sna}) for the soft nonfactorizable amplitude of
$\pi^- p \to \gamma^* n$ 
%for the exclusive Drell-Yan 
as a function of the Borel parameter $M_B^2$,
with 
%$Q'^2=5$~GeV$^2$, 
$|t|= 0.2$~GeV$^2$
%, and $\tau=0.2$
(left) 
and the corresponding cross section
%of
%$\pi^- p \to \gamma^* n$ 
as a function of $|t|$ (right),
%for $\tau = 0.2$ , 
with $Q'^2=5$~GeV$^2$and $\tau=0.2$
using the BMP2001 input
for the nucleon GPDs and $q_{\rm th}^2=0.7$~GeV$^2$ for the threshold parameter.
The solid and dashed curves are obtained using 
(\ref{sna}) 
and (\ref{eq_Hdu}), respectively.
    }
\label{fig:4}
%\end{center}
\end{figure}

Compared with the QCD factorization formula (\ref{eq_Hdu}), the pion DA does not appear in (\ref{sna}); instead,
% of the pion DA, 
we have the exponential factor, $\exp(- [(x - \xi)Q'^2]/[(x + \xi )M_B^2])$,
depending on the Borel parameter $M_B$,
characteristic of the QCD sum rule approach;
the relevant nonperturbative effects arising in the ``sum rule'' are encoded in the light-cone dominated quantities, the GPDs, 
and thus 
(\ref{sna}) corresponds to the light-cone sum rule 
for the nonfactorizable amplitude in the exclusive Drell-Yan process.
We note that the light-cone QCD sum rules have been derived for e.g., the pion electromagnetic form factor 
in \cite{Braun:1999uj}.
We present the behaviors of the soft nonfactorizable amplitude
%for the exclusive Drell-Yan 
from the light-cone sum rule (\ref{sna}), using the BMP2001 parameterization for the GPDs, 
which was used in the estimate shown in Fig.~\ref{fig:edycrossPLB}.
We note that the predictions using the QCD sum rules should not depend strongly on the Borel parameter, $M_B$,
introduced auxiliarily for the matching procedure.
In Fig.~\ref{fig:4}, we show (\ref{sna}) 
%for the exclusive Drell-Yan 
as a function of  $M_B^2$
with $Q'^2=5$~GeV$^2$, $|t|= 0.2$~GeV$^2$, and $\tau=0.2$.
We obtain good stability in a relevant range for $M_B^2$.
This result for (\ref{sna}) leads to the prediction to the cross section, $d\sigma_L/(dt dQ'^2)$,
due to the soft nonfactorizable mechanism for the exclusive pion-induced Drell-Yan process, $\pi^- p \to \gamma^* n$,
as shown by the solid curve in the right figure of Fig.~\ref{fig:4}
as a function of $t$, for the case with the same kinematics as in Fig.~\ref{fig:edycrossPLB}.
%the same result as a function of $\tau$ is also shown.
For comparison, we also plot 
%the cross section based on the LO QCD factorization formula by 
the dashed curve in Fig.~\ref{fig:4}, which is same as the black curve 
in Fig.~\ref{fig:edycrossPLB}.
The soft nonfactorizable mechanism gives the cross section larger by a factor of $\sim 5$ than the QCD factorization,
reflecting the 
$O(\alpha_s^0)$ and $O(\alpha_s^2)$ cross sections using (\ref{sna}) and  (\ref{eq_Hdu}), respectively,
and also shows the stronger dependence on $t$. The $\tau$ dependence of the soft nonfactorizable mechanism is also obtained~\cite{Tanaka:2017rym}. Our results indicate that the soft nonfactorizable mechanism 
should be very important at the J-PARC kinematics.
We note that even larger enhancement of the exclusive Drell-Yan cross section at J-PARC,
caused by a different type of soft mechanisms beyond the QCD factorization,
has been obtained by Goloskokov and Kroll~\cite{Goloskokov:2015zsa} (see also the discussion in \cite{Sawada:2016mao,Tanaka:2017rym}). 
Further study is needed to 
clarify the interplay in the soft/hard QCD mechanisms relevant for the J-PARC processes.

\noindent
{\bf Acknowledgments:} I thank 
T.~Sawada, W.~C.~Chang, S.~Kumano, J.~C.~Peng, and S.~Sawada
for helpful discussions and 
collaboration in \cite{Sawada:2016mao}.
% on which this work is based. 
I thank H.~Kawamura and P.~Kroll for useful discussions.
This work was supported by JSPS KAKENHI Grant Numbers JP25610058 and JP26287040.

\bibliographystyle{JHEP}
\bibliography{../bibdata/bibdatabase}% Produces the bibliography via BibTeX.
%\begin{thebibliography}{99}
%\bibitem{...}
%....
%
%\end{thebibliography}

\end{document}